\documentclass[11pt]{article}
\usepackage{amsmath}
\usepackage{graphicx}
\usepackage{natbib}
\usepackage{url} % not crucial - just used below for the URL

\usepackage[ruled,vlined,noend]{algorithm2e}
\SetAlFnt{\small}
\usepackage{amsfonts}
\usepackage{amsmath}
\usepackage{bbm}
\usepackage{floatrow}
\usepackage{subcaption}
\usepackage{hyperref}

\newcommand{\EE}{\mathrm{E}}

\def\T{ {\mathrm{\scriptscriptstyle T}} }

\def\diag{\mathrm{diag}}

\usepackage{amsmath}
\DeclareMathOperator*{\argmax}{argmax}
\begin{document}

\def\spacingset#1{\renewcommand{\baselinestretch}%
{#1}\small\normalsize} \spacingset{1.5}

%%%%%%%%%%%%%%%%%%%%%%%%%%%%%%%%%%%%%%%%%%%%%%%%%%%%%%%%%%%%%%%%%%%%%%%%%%%%%%

\begin{titlepage}

\begin{center}
{\Large Imputation Maximization Stochastic Approximation with
Application to Generalized Linear Mixed Models}

\vspace{.1in} Zexi Song\footnotemark[1] \& Zhiqiang Tan\footnotemark[1]

\vspace{.1in}
\today
\end{center}

\footnotetext[1]{Department of Statistics, Rutgers University. Address: 110 Frelinghuysen Road,
Piscataway, NJ 08854. E-mails: zexisong@stat.rutgers.edu, ztan@stat.rutgers.edu.}

\paragraph{Abstract.}

Generalized linear mixed models are useful in studying hierarchical data with possibly non-Gaussian responses. However, the intractability of
likelihood functions poses challenges for estimation. We develop a new method suitable for this problem, called imputation maximization stochastic approximation (IMSA).
For each iteration, IMSA first imputes latent variables/random effects, then maximizes over the complete data likelihood, and finally moves the
estimate towards the new maximizer while preserving a proportion of the previous value. %By design, IMSA is self consisent.
The limiting point of IMSA satisfies a self-consistency property and can be less biased in finite samples than
the maximum likelihood estimator solved by score-equation based stochastic approximation (ScoreSA).
Numerically, IMSA can also be advantageous over ScoreSA in achieving more stable convergence and
respecting the parameter ranges under various transformations such as nonnegative variance components.
This is corroborated through our simulation
studies where IMSA consistently outperforms ScoreSA.

%\vspace{-.2in}
\paragraph{Key words and phrases.} Stochastic approximation; Generalized linear mixed model; Latent variable model; Self consistency; Expectation maximization.

\end{titlepage}

\newpage
\spacingset{1.4} % DON'T change the spacing!

%-------------------------------------------------------------------
%-------------------------------------------------------------------
\section{Introduction}
\label{paper-3-sec:intro}

Latent variable modeling involves both latent and observed variables
such that their joint density is analytical tractable, but the marginal density of the
observed variable is analytically intractable.
Fitting latent variable models is known to be challenging due to analytically intractable likelihood functions.
In particular, an important class of latent variable models is generalized linear mixed models (GLMMs) \citep{McCulloch2008}
or hierarchical generalized linear models \citep{Gelman2007}.
Common existing methods for fitting GLMMs are hindered by various limitations including: requirement of a prior distribution for the parameters in the Bayesian approach;
potential bias caused by analytical approximation; high computational cost or sensitive and difficult tuning in current methods using expectation maximization (EM) or
stochastic approximation (SA) for likelihood inference. See Section \ref{paper-3-subsec:existing} for further discussion.
In this paper, we propose a new method called imputation maximization stochastic approximation, or IMSA for short,
and present simulation studies in support of the superior performance of IMSA.

The remainder of the paper is organized as follows. In Section~\ref{paper-3-sec:background} we introduce GLMMs in the broader context of latent variable models and review existing
methods; in Section~\ref{paper-3-sec:prop} we derive the proposed method, discuss its theoretical properties and compare it with current SA methods; in Section~\ref{paper-3-sec:sim} we present simulation studies on
two normal-logistic mixed models; in Section \ref{paper-3-sec:conclusion} we provide a conclusion.
%-------------------------------------------------------------------
%-------------------------------------------------------------------
%-------------------------------------------------------------------
\section{Background}
\label{paper-3-sec:background}
\vspace{-.075in}
%-------------------------------------------------------------------
\subsection{Latent variables models}
\label{paper-3-subsec:latent-variable}

Latent variable models are widely used to deal with complex data by introducing latent or hidden variables.
Formally, a latent variable model can be defined as follows. Let $y$ be observed variables and $u$ be latent variables
such that the joint density function $f(y,u;\theta)$ is analytically tractable in terms of $(y,u,\theta)$, where $\theta$ is a parameter vector. Typically, $f(y,u;\theta)$
is obtained as $f(y,u;\theta) = f(y|u;\theta)f(u;\theta)$, with both the marginal density $f(u;\theta)$ and the conditional density $f(y|u;\theta)$ analytically tractable.
However, the marginal density of $y$, defined in the integral form $f(y;\theta) = \int f(y, u;\theta)\,d u$, is analytically intractable. Since $f(y;\theta)$ is
also the likelihood function of parameter $\theta$ for fixed data $y$, its intractability creates difficulties (at least numerically) to likelihood and Bayesian methods alike.

%-------------------------------------------------------------------
\subsection{Generalized linear mixed models}
\label{paper-3-subsec:glmm}

For concreteness, we focus on a specific class of latent variable models called generalized linear mixed models (GLMMs), although our proposed method in Section \ref{paper-3-sec:prop} is applicable to general
latent variable models. GLMMs are a natural extension from generalized linear models by incorporating random effects. They can also be obtained from
linear mixed models by expanding the distribution of the response variables from Gaussian to an exponential family. There is an extensive literature about GLMMs \citep[e.g.,][]{McCulloch2008,Gelman2007}. Here we provide a brief description of GLMMs. Let $y = (y_1, ..., y_n)^{\T}$ be observed response variables, and $u = (u_1,...,u_q)^{\T}$
be latent variables (also referred to as random effects in this context). We introduce covariates $x_i$ and $z_i$ for $i = 1,...n$, associated with $p$-dimensional fixed effects $\beta$ and
random effects $u$ respectively. Then conditional on $u$, the response variables $(y_1,\ldots,y_n)$ are independent and each $y_i$ is distributed with its density $f(y_i|u;\beta)$ in an exponential family. Through a
link function $g$, the conditional mean satisfies $\EE[y_i|u] = g^{-1}(x_i^{\T} \beta + z_i^{\T} u)$. Finally a GLMM is fully specified with the marginal density $f(u)$, commonly
chosen to be Gaussian.

In subsequent discussion, we mainly consider GLMMs with a binary response $y$, a logit link function, and normally distributed $u$. The resulting logistic-normal mixed
model is defined as follows:
\begin{align}
    & P(y_i = 1|u) = \text{expit}(x_i^{\T} \beta + z_i^{\T} u), \quad i = 1,...,n, \label{paper-3-eq:model1}\\
    & u\sim N(0,\diag(\sigma^2)),\quad \sigma^2 = (\underbrace{\sigma_1^2,...,\sigma_1^2}_{q_1},\, \underbrace{\sigma_2^2,...,\sigma_2^2}_{q_2}, ... , \underbrace{\sigma_K^2,...,\sigma_K^2}_{q_K} ),\label{paper-3-eq:model2}
\end{align}
where $\sum_{k = 1}^K q_k = q$. In the formulation above we allow for $K$ different groups of latent variables each with
variance $\sigma_k^2$ and group size $q_k$. The objective is then to estimate the parameter vector $\theta = (\beta^{\T},(\sigma^2)^{\T})^{\T}$. Define the linear predictor $\eta = (\eta_1,\eta_2,...,\eta_n)^{\T}$ with $\eta_i = x_i^{\T} \beta + z_i^{\T} u$. Then according to (\ref{paper-3-eq:model1}) and (\ref{paper-3-eq:model2}), the joint (complete-data) density and log-likelihood are given in a closed-form as follows:
\begin{align}
    & f(y,u;\theta) = \exp\left\{\sum_{i = 1}^n \left[y_i\eta_i - \log(1+e^{\eta_i})\right]\right\}\cdot \exp\left\{- \frac{1}{2} u^{\T} [\diag(\sigma^2)]^{-1} u \right\} \cdot |Det(\diag(\sigma^2))|^{-\frac{1}{2}} , \label{paper-3-eq:jointlik} \\
   &\log f(y,u;\theta) = \sum_{i = 1}^n \left[y_i\eta_i - \log(1+e^{\eta_i})\right] - \frac{1}{2} u^{\T} [\diag(\sigma^2)]^{-1} u - \frac{1}{2}\sum_{k = 1}^K q_k \log \sigma_k^2.\label{paper-3-eq:jointloglik}
\end{align}

%-------------------------------------------------------------------
\subsection{Existing methods}
\label{paper-3-subsec:existing}

We briefly discuss existing methods for fitting GLMMs. Bayesian inference has been extensively studied, notably through various posterior sampling methods; see for example
\citet{Gamerman1997} and \cite{Gelman2008}. Such development is facilitated by the availability of a joint distribution
for $(u,\theta|y)$ once a prior distribution on $\theta$ is introduced. The posterior credible intervals for fixed parameters and posterior predictive intervals for
latent variables are operationally appealing to practitioners, although suitable frequentist justification remains to be fully developed \citep{Jiang2013, Chae2019}.
From a methodological perspective, the necessity of a prior distribution may also be a limitation for the Bayesian approach, compared with likelihood based methods
which are prior-free and can be used for similar purposes.

Existing likelihood based methods broadly fall into three categories. The first type is Laplace and related analytical approximation, such as penalized quasi-likelihood
\citep{Schall1991,Breslow1993} and H-likelihood method \citep{Lee2006} among others. %and integration over Taylor expansions of the joint density \citep{Goldstein1991}.
While these methods are convenient, they are also potentially biased due to the approximate nature.

The second type includes expectation-maximization (EM) \citep{Dempster1977} and its variants. For
GLMMs, the expectation required for the E-step in EM is analytically intractable. By Monte Carlo EM (MCEM),
the intractable expectation is approximated via Monte Carlo samples \citep{Wei1990, McCulloch1997, Booth1999, Fort2003}.
This procedure is computationally intense and the tuning of its hyper-parameters can be difficult.
Alternatively, stochastic EM (StEM) \citep{Celeux1986, Nielsen2000} alleviates the computational burden of MCEM by sampling a single copy of latent variables $u$ at each update, instead of using multiple copies. We
henceforth refer to StEM as imputation-maximization (IM) and describe an extension as follows, which in part motivates our proposed method.

\vspace{.05in}\textit{Imputation-maximization} (IM). \quad Given initial values $\tilde{\theta}_0$ and $u_0$, iterate for $t = 1,2, ... $, \vspace{-.1in}
\begin{itemize}\addtolength{\itemsep}{-.1in}
    \item Sample $u_t$ by MCMC given $u_{t-1}$ leaving $ f(u|y;\tilde{\theta}_{t-1})$ invariant. \vspace{-.05in}
    \item Compute $\tilde{\theta}_t = \argmax_{\theta}\log f(y,u_t;\theta)$ \vspace{-.05in}.
\end{itemize}
In fact, the IM algorithm has been studied in \citet{Nielsen2000}, assuming that $u_t$ is exactly drawn from $f(u|y;\tilde\theta_{t-1})$, independently of $u_{t-1}$.
For GLMMs, exact sampling from $f(u|y;\theta)$ is infeasible and a direct extension based on Markov chain Monte Carlo (MCMC) can be used.
Under suitable conditions, the IM sequence $(\tilde{\theta}_0, \tilde{\theta}_1, \tilde{\theta}_2, ...)$
may be expected to converge to a non-degenerate distribution, hence randomly fluctuating instead of converging to a fixed point.
A point estimate can be formed by taking the average of $\{\tilde{\theta}_t\}$ up to some large $t$ (possibly after some burn-in iterations).
However, as we find from numerical experiments,
IM is prone to getting stuck near zero over long periods in estimating the variance component, i.e. $\sigma^2$ in (\ref{paper-3-eq:model2}).
This drawback may be caused by the fact that $\tilde{\theta}_{t-1}$ is completely refreshed by a new maximizer in each iteration.

For the remainder of this section, we discuss the third type of likelihood methods: stochastic approximation (SA), which is also exploited by our proposed method.
Pioneered by \citet{Robbins1951}, SA can be
interpreted as a root finding algorithm. Consider a function $h(\theta)$ that cannot be directly evaluated. Instead, only a stochastic version $H(u;\theta)$ is available
such that $\EE_\theta [ H(u; \theta)] = h(\theta)$, where $\EE_{\theta}[\cdot]$ denotes the expectation for $u$ under a probability density function $p(\cdot; \theta)$.
Then SA finds a solution $\theta^*$ to the equation $h(\theta) = 0$ by the following.

\vspace{.05in}\textit{Stochastic approximation} (SA). \quad Given initial values $\tilde{\theta}_0$ and $u_0$, iterate for $t = 1,2, ... $, \vspace{-.1in}
\begin{itemize}\addtolength{\itemsep}{-.1in}
    \item Sample $u_t$ by MCMC given $u_{t-1}$ leaving $p(u; \tilde{\theta}_{t-1})$ invariant. \vspace{-.05in}
    \item Update $\tilde{\theta}_t = \tilde{\theta}_{t-1} + \gamma_t H(u_t; \tilde{\theta}_{t-1})$, \vspace{-.05in}
\end{itemize}
where $\gamma_t$ is a sequence of step sizes.
For the classical SA in \cite{Robbins1951}, $u_t$ is exactly drawn from $p(u;\tilde\theta_{t-1})$, independently of $u_{t-1}$.
In the description above, $u_t$ is sampled by MCMC depending on both $\tilde\theta_{t-1}$ and $u_{t-1}$.
Under regularity conditions, if the step sizes are chosen to be $\gamma_t = 1/t$, then $\tilde{\theta}_t$ can be shown to converge to $\theta^*$ almost surely
\citep[e.g.,][]{Benveniste1990,Chen2002}.
The SA framework is very flexible and can accommodate a wide range of problems.
In order to apply SA to GLMMs, suitable functions $h$ and $H$, and probability density $p(u;\theta)$ need to be selected.
For maximization of the marginal likelihood $f(y;\theta)$,
a standard choice of $h$  is $h(\theta) = \frac{\partial}{\partial \theta}\log f(y;\theta)$,
with the corresponding $H(u;\theta) = \frac{\partial}{\partial \theta}\log f(y, u;\theta)$ and $p(u; \theta) = f(u|y;\theta)$,
where  $\EE_{\theta}[\cdot]$ denotes the expectation  with respect to latent variables $u\sim f(u|y;\theta)$.
Then it is straightforward to verify the SA condition $\EE_{\theta}[H(u;\theta)] = h(\theta)$. Fitting GLMMs with this type of SA is studied by \citet{Gu1998} and \citet{Zhu2002}. We refer to it as ScoreSA
since it aims to solve the score equation $\frac{\partial}{\partial \theta}\log f(y;\theta) = 0$, and formally define it in Algorithm~\ref{paper-3-alg:ScoreSA}.

\begin{algorithm}[!tph] %[H]
    \SetAlgoNoLine
    Initialize $\hat{\theta}_0$ and $u_0$\\
    \For{ $t = 0,1,2,..., T$}{
        Sample $u_t$ by MCMC given $u_{t-1}$  leaving $ f(u|y;\hat{\theta}_{t-1})$ invariant \\
        Update $\hat{\theta}_t = \hat{\theta}_{t-1} + \gamma_t\left\{ \frac{\partial}{\partial \theta}\log f(y, u_t;\theta)\mid_{\theta = \hat{\theta}_{t-1}}\right\}$ \quad with $\gamma_t$ a step size
    }
    Output $\hat{\theta}_T$
    \caption{ScoreSA}\label{paper-3-alg:ScoreSA}
\end{algorithm}

The limiting point that ScoreSA converges to is the marginal MLE $\hat{\theta}_{ML} = \argmax_{\theta}\log f(y;\theta)$. The algorithms in
\citet{Gu1998} and \citet{Zhu2002}
are more sophisticated with additional features and the associated tuning such as approximation of the hessian matrix of $\log f(y, u;\theta)$. Nevertheless, Algorithm~\ref{paper-3-alg:ScoreSA}
conveys the main ideas and may perform adequately subject to careful tuning.
For ScoreSA, the magnitude of step size $\gamma_t$ has no simple interpretation (compared with IMSA in Section \ref{paper-3-sec:prop})
thus making the tuning challenging. As we illustrate in Section~\ref{paper-3-sec:sim},
the performance of ScoreSA can be sensitive to the choice of $\gamma_t$. Another disadvantage of ScoreSA is that the gradient $\frac{\partial}{\partial \theta}\log f(y, u;\theta)$
can vary drastically for $\theta$ defined on different parameter scales (e.g. original v.s. log scale for the variance components).
To achieve reasonable performance, a suitable parameter scale often needs to be selected for parameter transformation when applying ScoreSA.

%-------------------------------------------------------------------
%-------------------------------------------------------------------
%-------------------------------------------------------------------
\section{Proposed method}
\label{paper-3-sec:prop}
Motivated by the discussion in Section \ref{paper-3-subsec:existing}, we seek to develop a method that is prior-free, computationally efficient, easy to tune and also
accommodates flexible parameter transformations, e.g., with variance components in the original or log scale.
To achieve this, we build upon the IM algorithm. As mentioned in Section~\ref{paper-3-sec:background} , the fluctuating behavior
of IM can be problematic. Therefore instead of completely replacing $\tilde{\theta}_{t-1}$, we add an SA type update after the maximization step, such that only a portion
of $\tilde{\theta}_{t-1}$ is replaced by the maximizer, controlled through a learning rate or step size $\gamma_t$. This yields our proposed method, imputation maximization
stochastic approximation (IMSA) as described in Algorithm~\ref{paper-3-alg:IMSA}

\begin{algorithm}[!tph] %[H]
    \SetAlgoNoLine
    Initialize $\tilde{\theta}_0$ and $u_0$\\
    \For{ $t = 0,1,2,..., T$}{
        Sample $u_t$ by MCMC given $u_{t-1}$ leaving $ f(u|y;\tilde{\theta}_{t-1})$ invariant  \qquad\#Imputation \\
        Compute $\tilde{\theta}_{t-1/2} = \argmax_{\theta}\log f(y,u_t;\theta)$  \qquad\qquad\qquad\qquad\qquad\quad\,     \#Maximization \\
        Update $\tilde{\theta}_t = \tilde{\theta}_{t-1} + \gamma_t(\tilde{\theta}_{t-1/2} - \tilde{\theta}_{t-1})$ \quad with $\gamma_t$ a step size \,\,\,\,\,\quad\quad\quad\#Update/Shrinkage
    }
    Output $\tilde{\theta}_T$
    \caption{IMSA}\label{paper-3-alg:IMSA}
\end{algorithm}

Our formulation of IMSA is not restricted to GLMMs and can potentially handle general latent variable models discussed in Section~\ref{paper-3-subsec:latent-variable}. The key consideration
for practicality is whether the maximization step is easily implementable. In the GLMM case, the maximization can
be carried out efficiently. In (\ref{paper-3-eq:jointloglik}), the terms containing $\beta$ are
\[
    \sum_{i = 1}^n \left[y_i (x_i^{\T}\beta) - \log(1+e^{\eta_i})\right].
\]
Therefore maximizing $\log f(y,u;\theta)$ over $\beta$ given $u$ is equivalent to find the MLE for a GLM with offset \citep{McCullagh1989}. Here the offset values are $z_i^{\T} u$. This problem
is well studied with fast algorithms readily available.  For the experiments in Section~\ref{paper-3-sec:sim} we use Python module \texttt{statsmodel.api.GLM} which implements iteratively reweighted least squares.
For $\sigma^2$, we denote as $u^{(k)}$ the sub-vector of $u$ corresponding to the $k$-th variance group. Then the derivative with respect to $\sigma^2_k$ is
\[
    \frac{\partial}{\partial \sigma^2_k} \log f(y,u;\theta) = \frac{1}{2}\sigma_k^{-4}[u^{(k)}]^{\T}u^{(k)} - \frac{1}{2}q_k\sigma_k^{-2}.
\]
Setting the above to zero, we obtain the maximizer $\tilde{\sigma}^2_k = [u^{(k)}]^{\T}u^{(k)}/q_k$, which can be understood as the sample (zero-centered) variance of latent variables in the $k$-th group.

IMSA can be formally put in the SA framework as follows.
By matching the update from $\tilde\theta_{t-1}$ to $\tilde\theta_t$ in Algorithm \ref{paper-3-alg:IMSA} with the general SA update in Section \ref{paper-3-subsec:existing},
the stochastic function associated with IMSA is
\[
H(u; \theta) = M(u;\theta) - \theta,\quad \text{ where }\quad M(u;\theta) = \textstyle \argmax_{\theta}\log f(y,u;\theta).
\]
Then by construction, the corresponding $h$ function is $h(\theta) = \EE_{\theta}[M(u;\theta)] - \theta$. Again $\EE_{\theta}$ stresses that the
expectation is taken over $f(u|y;\theta)$. Setting $h(\theta) = 0$, the limiting point of IMSA denoted by $\tilde{\theta}_{IMSA}$, satisfies the following,
\begin{equation}
    \label{paper-3-eq:selfcon}
    \theta = \EE_{\theta}[M(u;\theta)].
\end{equation}
Equation (\ref{paper-3-eq:selfcon}) is called the self-consistency condition \citep{Meng2007}. Self-consistency principle states that: a desired parameter estimate $\tilde{\theta}$
should equal to the expectation of possible parameter estimates obtained from randomly imputed data according to $\tilde{\theta}$ (in conjunction with observed data $y$). This
can be seen to provide a statistical justification for IMSA.

In the following, we compare IMSA and ScoreSA in several ways.
First, the IMSA estimator $\tilde{\theta}_{IMSA}$ is in general numerically distinct from the MLE $\hat{\theta}_{ML}$ solved by ScoreSA.
Under standard regularity conditions, however, they can be shown to be asymptotically equivalent to each other
\citep{Meng2007}:
\[
    \tilde{\theta}_{\text{IMSA}} = \hat{\theta}_{\text{ML}} + o_p(n^{-\frac{1}{2}}) .
\]
In fact, by asymptotic expansion in the complete-data model we have in the IMSA update,
\[
    \tilde{\theta}_{t - 1/2} - \tilde{\theta}_{t - 1} = -\left\{\frac{\partial^2}{\partial \theta \partial \theta^{\T}} \log f(y, u_t;\theta)\mid_{\theta = \tilde{\theta}_{t-1}}\right\}^{-1}\left\{\frac{\partial}{\partial \theta} \log f(y, u_t;\theta)\mid_{\theta = \tilde{\theta}_{t-1}}\right\} + o_p(n^{-1/2}).
\]
Hence the leading term in $\tilde{\theta}_{t - 1/2} - \tilde{\theta}_{t - 1}$ is the gradient of $\log f(y, u_t;\theta)$ with its negative inverse hessian matrix multiplied in front.
Contrasting this with the ScoreSA update
\[
 \hat{\theta}_t = \hat{\theta}_{t-1} + \gamma_t\left\{ \frac{\partial}{\partial \theta}\log f(y, u_t;\theta)\mid_{\theta = \hat{\theta}_{t-1}}\right\},
\]
we see that the IMSA update is a preconditioned version of the ScoreSA update. Preconditioning is analogous to the use of Newton-Raphson as opposed to gradient descent for
achieving faster convergence \citep[e.g.][]{Girolami2011}. Therefore, we expect that IMSA will generally enjoy faster and more stable convergence than
ScoreSA, even though the limits of these algorithms may differ on a fixed dataset. In addition, in small or moderately-sized samples, $\tilde{\theta}_{IMSA}$ can be less biased
than $\hat{\theta}_{ML}$ by related analysis in \cite{Fang2018}.

Second, the IMSA update $\tilde{\theta}_t = \tilde{\theta}_{t-1} + \gamma_t(\tilde{\theta}_{t-1/2} - \tilde{\theta}_{t-1})$ can be viewed as a shrinkage update, related to the IM update $\tilde{\theta}_{t-1/2}$.
The algorithm moves from $\tilde{\theta}_{t-1}$ towards $\tilde{\theta}_{t-1/2}$ while maintaining $\tilde{\theta}_{t-1}$ by the amount of $1 - \gamma_t$.
Thus the learning rate $\gamma_t$ in IMSA has a clear interpretation as the proportion by which the maximizer obtained from newly imputed data $\tilde{\theta}_{t-1/2}$ will
be incorporated. The previously mentioned IM (Section~\ref{paper-3-subsec:existing}) corresponds to the special case where $\gamma_t \equiv 1$. Because each $\tilde{\theta}_{t-1/2}$ is already
in the proper range of $\theta$, IMSA estimates $\{\tilde{\theta}_{t}\}$ remain in the proper range for a convex parameter space. In this sense, IMSA is range-respecting and safeguards
against invalid parameter values (e.g., a negative variance estimate), regardless of the scale that $\theta$ is specified on. In contrast, when estimating $\sigma^2$ directly with ScoreSA,
it is possible to obtain a negative value since the magnitude of the gradient does not preserve the range of $\sigma^2 \geq 0$. Thus for ScoreSA, it is customary to update the variance on
$\log\sigma$ scale, whereas IMSA is not bound by this constraint. One subtlety is that when applying a transformation of $\theta$, while the maximization step is invariant, the shrinkage step
will lead to different results, as illustrated in Section~\ref{paper-3-sec:sim}.

Third, compared with ScoreSA, IMSA involves a higher computational cost per iteration due to the maximization of the log-likelihood $\log f(y,u_t;\theta)$ with
imputed data $u_t$. However, the cost increase is limited in the settings where fast algorithms are available for maximizing $\log f(y,u_t;\theta)$,
such as in generalized linear mixed models.
Moreover, the majority of computational cost per iteration in IMSA as well as ScoreSA is often incurred by MCMC sampling to impute the latent variable $u_t$.
Hence the cost per iteration from IMSA may be only slightly higher than ScoreSA,
while IMSA tends to achieve more stable convergence (with minimal tuning) and more accurate estimation,
as shown in our numerical experiments.

As a side note, the variance matrix of $\tilde{\theta}_{IMSA}$ can be estimated by exploiting the missing information identity in \citet{Louis1982}. Combined with online variance
formulas \citep{Welford1962}, this calculation can be completed in a single pass of IMSA, hence no separate simulation is needed. Furthermore, the variance estimators
can also be used to construct a preconditioned IMSA algorithm that is asymptotically optimal by SA theory, similarly as in \citet{Gu1998} and \citet{Zhu2002}. Details are given by
Algorithm~\ref{paper-3-alg:IMSA3} in the Appendix. Nevertheless the actual implementation of variance estimation requires complicated tuning and converge monitoring with multiple step sizes.
Therefore the experiments in the following section focus solely on the point estimate of $\theta$.

%-------------------------------------------------------------------
%-------------------------------------------------------------------
%-------------------------------------------------------------------
\section{Simulation studies}
\label{paper-3-sec:sim}

We compare IMSA with ScoreSA through two logistic-normal mixed models. We include both IMSA that updates variance on the original $\sigma^2$ scale and IMSA
that updates variance on the $\log\sigma$ scale. We label the latter as IMSA-log. For ScoreSA, we always update variance on the $\log\sigma$ scale. This decision is due to
two reasons: first as previously mentioned, ScoreSA cannot guarantee the positivity of $\sigma^2$ when updating on the original scale; secondly during trial runs,
ScoreSA frequently encounters gradient explosion and breaks down when updating on the original scale. To emphasize the point that our proposed method can achieve good results with minimal
amount of tuning, we simply set the learning rate $\gamma_t = t^{-1}$ for both IMSA and IMSA-log. In contrast, ScoreSA is sensitive to the choice of $\gamma_t$ and requires
careful tuning for stable and reasonable performance. Therefore we choose a list of different $\gamma_t$'s for ScoreSA taking the form of $\gamma_t = \min(t^{-1},t^{-1}_0)$ with
the constant $t_0\in\{1,5,10,25,50,100\}$. We label these different versions of the algorithm as ScoreSA-1/2/3/4/5/6 corresponding to (in the increasing order) the six $t_0$ values
with ScoreSA-1 for $t_0 = 1$, ScoreSA-2 for $t_0 = 5$, etc. To sample from the intractable distribution $f(u|y;\theta)$, in the simulations we employ multiple MCMC samplers (in parallel).
Specifically, we use preconditioned Metropolis-adjusted Langevin algorithm (pMALA) \citep{Besag1994,Roberts1996}, shown as Algorithm~\ref{paper-3-alg:pMALA} in the Appendix.
Further simulation details and additional results are also provided in the Appendix.
All algorithms are implemented in Python. In both simulation settings, each individual run
is carried out using an Intel Skylake CPU with two cores and $2000$ megabytes of memory.

%-------------------------------------------------------------------

\subsection{Booth--Hobert example}
\label{paper-3-subsec:simlog}

Consider a simple logistic-normal mixed model that is studied in \citet{Booth1999}: for $i = 1,...,10$ and $j = 1, ... ,15$, \vspace{-.05in}
\begin{align}
   & y_{ij}|u_i\text{ are independent Bernoulli variables with } P(y_{ij} = 1|u_i) = \text{expit}(\beta x_{ij} + u_i),\label{paper-3-eq:boothexample} \\
   & x_{ij} = \frac{j}{15},\quad u_i\overset{i.i.d.}{\sim} N(0,\sigma^2).\nonumber
\end{align}
This is a simplified version of the original model discussed in \citet{McCulloch1997} and we refer to it as Booth--Hobert example. We use the reported true parameter values
$\beta = 5$ and $\sigma^2 = 0.5$ as in \citet{Booth1999} and randomly generate $100$ sets of data from (\ref{paper-3-eq:boothexample}). Then for every data set, we estimate $(\beta,\sigma^2)$
with IMSA, IMSA-log, and ScoreSA-1/2/3/4/5/6. Each algorithm is run for $2000$ iterations.
On average, an individual run takes $67$ seconds for IMSA and IMSA-log, and $37$ seconds for ScoreSA-1/2/3/4/5/6.
The initial values are dispersed (uniformly at random) over the
intervals $\beta\in(1,2), \sigma^2\in(0.5,1.5)$ for each data set.

\begin{figure}[t]
    \centering
    \includegraphics[width = 0.95\textwidth]{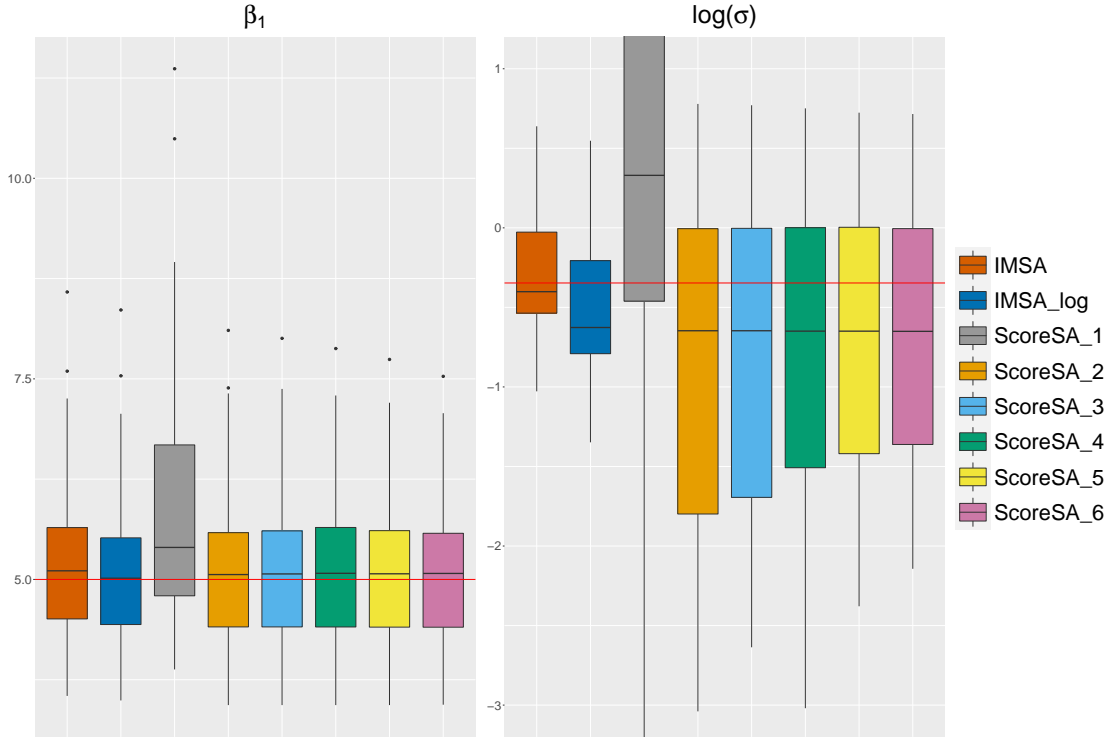}
    \caption{Boxplots of parameter estimates in Booth--Hobert example. Red lines mark true values.}
    \label{paper-3-fig:logitbox}
\end{figure}

Boxplots of the estimates are presented in Figure~\ref{paper-3-fig:logitbox} with ground truth marked by red horizontal lines. Notice that the variance component is plotted
on the $\log \sigma$ scale, although IMSA updates on the original $\sigma^2$ scale. We see that for $\beta$, all methods except for ScoreSA-1 have comparable performance and
yield satisfactory results, whereas ScoreSA-1 overestimates $\beta$.
For the more challenging problem of estimating $\sigma^2$, IMSA clearly has the overall best performance followed by IMSA-log which shows similar spread but
underestimates compared to IMSA. ScoreSA-1 estimates are non-stable and outside of the plotting limits which encapsulates all other methods. From ScoreSA-2 to ScoreSA-6, the
spreads decrease as would be expected since these methods use progressively smaller initial $\gamma_t$. The six versions of ScoreSA show
comparable amount of underestimation to IMSA-log, but all have larger spreads indicating inconsistency over repeated runs.

%-------------------------------------------------------------------
\subsection{Salamander mating model}
\label{paper-3-subsec:simsal}

The second example we use is a model regarding the mating behavior among salamanders of two different species, originally investigated in \citet{McCullagh1989}.
Let $y_{ij}, i,j = 1,...,60$ be the indicator of mating between female $i$ and male $j$, with $y_{ij} = 1$ corresponding to a successful mating and $y_{ij} = 0$ otherwise.
Note that out of all possible female/male pairings, only $360$ are observed. We label the two species by A and B. The response $y_{ij}$ depends on both the species through
fixed effects and sex through random effects. Let $\beta = (\beta_1,\beta_2,\beta_3,\beta_4)^{\T}$ be fixed effects for the female/male species combination in
the order$(A/A,A/B,B/A,B/B)$, e.g., $\beta_2$ corresponds to a female from species A and a male from species B, etc. Let $x_{ij}$, a vector of length $4$, be the $(0,1)$ encoding of species combination
of the ($i,j$)-th pair. Finally we denote $u_i^{Female}$ and $u_j^{Male}$ as random effects from the $i$-th female and $j$-th male respectively. Then the mixed effect model can be written as
\begin{align}
    & y_{ij}|u_i^{Female}, u_j^{Male} \sim Bernoulli(\pi_{ij}), \text{ independently}\nonumber,\\
    & \pi_{ij} = \text{expit}(x_{ij}^{\T}\beta + u_i^{Female} + u_j^{Male} ), \label{paper-3-eq:sal}\\
    & u_i^{Female}\overset{i.i.d.}{\sim} N(0,\sigma_1^2), \quad u_j^{Male}\overset{i.i.d.}{\sim} N(0,\sigma_2^2).\nonumber
 \end{align}
With $\sigma^2 = (\sigma_1^2, \sigma_2^2)^{\T}$, the full parameter vector is $\theta = (\beta^{\T},\,  (\sigma^2)^{\T} )^{\T}$. \citet{Booth1999} report the MLE of the original data
to be $\beta_1 = 1.03,\beta_2 = 0.32, \beta_3 = -1.95,\beta_4 = 0.99,\sigma^2_1 = 1.4$ and $\sigma^2_2 = 1.25$. We use these values as data generating parameters and
create $100$ synthetic data sets from (\ref{paper-3-eq:sal}). We then estimate $\theta$ on each data set. Because this model is more complicated than Booth--Hobert example (now with
$6$ parameters and $120$ latent variables), all methods are run for $4000$ iterations.
On average, an individual run takes $475$ seconds for IMSA and IMSA-log, and $396$ seconds for ScoreSA-1/2/3/4/5/6.
Initial values are dispersed over the following intervals,
$\beta_1\in(0,2), \beta_2\in(-1,1), \beta_3\in(-3,-1), \beta_4\in(0,2), \sigma_1^2\in(1,2.5)$ and $\sigma_2^2\in(1,2.5)$.

\begin{figure}[t]
    \centering
    \includegraphics[width = 0.95\textwidth]{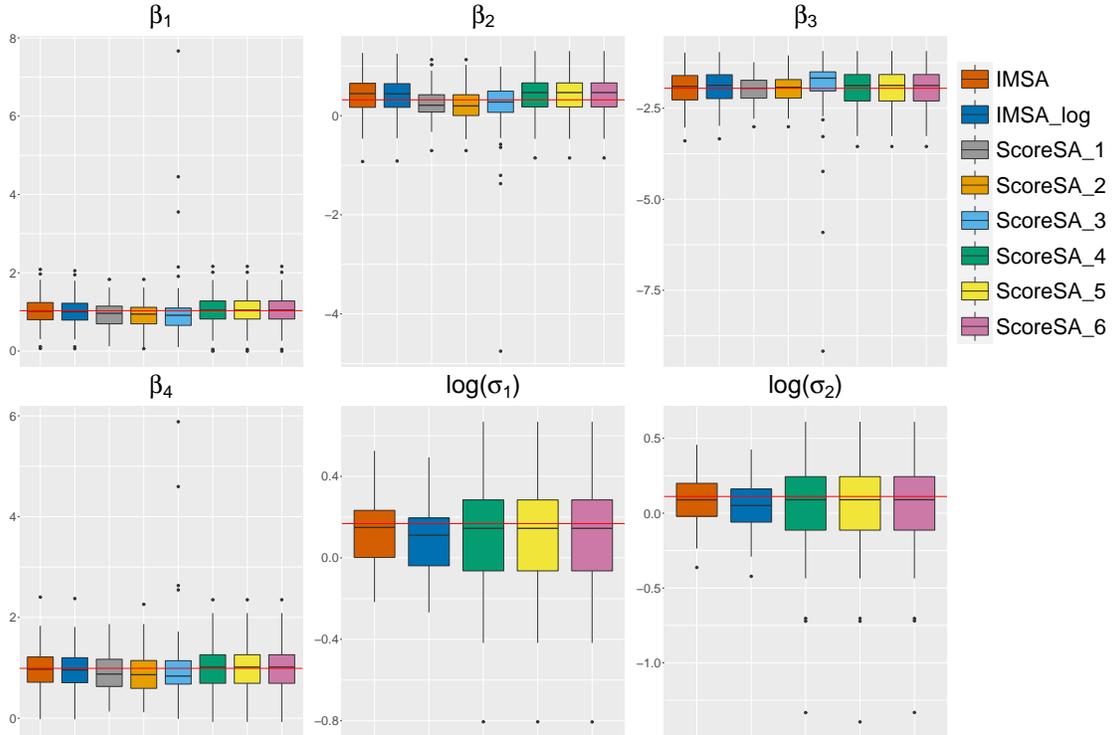}
    \caption{Boxplots of parameter estimates for the salamander mating model. Red lines mark true values. Variance estimates from ScoreSA-1/2/3 are omitted
    due to poor performance.}
    \label{paper-3-fig:salbox}
\end{figure}

The results are plotted in Figure~\ref{paper-3-fig:salbox}. ScoreSA-1/2/3 all encounter instability and produce extremely large values of $\sigma_1^2,\sigma_2^2$. In that sense
we consider that these three methods fail on the salamander model and exclude them from the plots of variance component. All remaining methods are highly comparable in
$\beta$ with reasonably good fits. For the more challenging problem of estimating variance components, however, IMSA is distinctly superior over the others in both $\sigma_1^2$ and $\sigma_2^2$. IMSA-log shows variation that is
comparable to IMSA. It appears to underestimate $\sigma_1^2$ and $\sigma_2^2$, but still covers the true values. For ScoreSA-4/5/6, their median estimates of $\log\sigma_1$ and $\log \sigma_2$
are slightly more accurate than IMSA-log, but they show wider spread and overall outperformed by IMSA.

%-------------------------------------------------------------------
%-------------------------------------------------------------------
%-------------------------------------------------------------------
\section{Conclusion}
\label{paper-3-sec:conclusion}
We develop imputation maximization stochastic approximation with application to generalized linear mixed models.
The proposed IMSA method is prior-free, computationally fast,
and easy to tune, even though the cost per iteration may be slightly higher than ScoreSA.
The method also allows flexible parameter transformations %enjoys the self-consistency property
and ensures that the corresponding estimates fall in the proper parameter ranges such as nonnegative variance components.
In two sets of numerical experiments, IMSA consistently outperforms its main competitor ScoreSA.
We also observe that IMSA yields better results when updating the variance components on the original scale than on the log scale. For future work, it is desired to investigate
in detail the estimation of the variance of $\tilde{\theta}_{IMSA}$. Moreover, it is interesting to extend beyond GLMMs, and apply IMSA to other suitable latent variable models.

\bibliographystyle{apalike}
\bibliography{imsaref}
\newpage
\section*{Appendix A: Simulation details}

Let $X_{n\times p}$ and $Z_{n\times q}$ be design matrices associated with covariates $x_{i}$ and $z_i$ respectively. The specific form of $X,Z$ in Booth--Hobert example and
salamander mating model are provided in our simulation codes which are available upon request. Define vectors of linear predictors as $\eta^{(1)} = X\beta, \eta^{(2)} = Zu$ so
that $\eta^{(1)} + \eta^{(2)} = \eta$. Transform the variance component as $\tau_k = \log \sigma_k$ for $k = 1, ..., K$. Then the gradient $\frac{\partial}{\partial \theta} \log f(y,u;\theta)$
required for ScoreSA is given by
\[
\frac{\partial}{\partial \beta} \log f(y,u;\theta) = X^{\T}(y - \text{expit}(\eta)), \quad \frac{\partial}{\partial \tau_k} \log f(y,u;\theta) = \exp(-2\tau_k)[u^{(k)}]^{\T}u^{(k)} - q_k.
\]
In the above, the expit function is understood to be applied component-wise to the vector $\eta$.

For sampling latent variables $u\sim f(u|y;\theta)$, it is convenient to work with the negative log-likelihood, or potential energy function,
$Q$ such that $\exp(-Q(u))\propto f(u|y;\theta)$. The expression of $Q$ is given by
\[
    Q(u) = \frac{1}{2}u^{\T}[\diag(\sigma^2))]^{-1}u - \sum_{i = 1}^n(y_i\eta_i^{(2)} - \log(1 + e^{\eta_i})).
\]
In the argument of $Q$, we suppress the dependence on $y$ and parameters $\theta$ as they are fixed during each imputation step. The gradient $\nabla Q$ and hessian $\nabla^2 Q$
are also needed for the imputation:
\[
  \nabla Q(u) = [\diag(\sigma^2))]^{-1}u - Z^{\T}(y - \text{expit}(\eta)),\quad \nabla^2 Q(u) = [\diag(\sigma^2))]^{-1} + Z^{\T}\diag[\text{expit}(\eta)(1 - \text{expit}(\eta))]Z.
\]

The implementation of ScoreSA and IMSA in Section~\ref{paper-3-sec:sim} are given by Algorithm~\ref{paper-3-alg:IMSA2} and \ref{paper-3-alg:ScoreSA2}, incorporating multiple imputation.
For the imputation (described in Algorithm~\ref{paper-3-alg:pMALA}), in both examples we use $4$ parallel MCMC samplers that run for $20$ steps in between each parameter update. Preconditioning for MCMC is only applied after $500$ iterations,
that is $T_0 = 500$. During the first $500$ iterations, the vanilla MALA is used. The sampling step size $\epsilon$ is periodically adjusted to maintain MCMC acceptance rate at around $60\%$.

\begin{algorithm}[!htp] %[H]
    \SetAlgoNoLine
    Initialize $\tilde{\theta}_0$ and $m$ copies of latent variables $u_{01},u_{02},...,u_{0m}$ \\
    \For{ $t = 0,1,2,..., T$}{
        \eIf{$t\leq T_0$}{
            \For{$j = 1,2, ..., m$}{
                Sample $u_{tj}$ given $(u_{(t-1)j},\tilde{\theta}_{t-1})$ leaving $f(u|y;\tilde{\theta}_{t-1})$ invariant by Algorithm~\ref{paper-3-alg:pMALA} \textbf{without preconditioning}
            }
           }{
            \For{$j = 1,2, ..., m$}{
                Sample $u_{tj}$ given $(u_{(t-1)j},\tilde{\theta}_{t-1})$ leaving $f(u|y;\tilde{\theta}_{t-1})$ invariant by Algorithm~\ref{paper-3-alg:pMALA} \textbf{with preconditioning}
            }
          }
          \For{$j = 1,2, ..., m$}{
            Compute $\beta^{\dagger}_{j}$ as MLE regression coefficients of $GLM(y,X)$ with offset $Z\, u_{tj}$\\
            \For{$k = 1,2, ..., K$}{
                Compute $(\sigma^{2})^{\dagger(k)}_{j} = [u^{(k)}_{j}]^{\T}u^{(k)}_j/q_k$
            }
            Set $(\sigma^{2})^{\dagger}_{j} = (\cdots(\sigma^{2})^{\dagger(k)}_{j} \cdots)$
          }
        \eIf{updating on the $\log \sigma$ scale}{
            Set $\tilde{\theta}_{t - 1/2} = \left(\frac{1}{m}\sum_{j=1}^m \beta^{\dagger}_{j},\frac{1}{2m}\sum_{j=1}^m \log[ (\sigma^{2})^{\dagger}_{j}]  \right)$
        }{
            Set $\tilde{\theta}_{t - 1/2} = \left(\frac{1}{m}\sum_{j=1}^m \beta^{\dagger}_{j},\frac{1}{m}\sum_{j=1}^m (\sigma^{2})^{\dagger}_{j}  \right)$
        }
        Update $\tilde{\theta}_t = \tilde{\theta}_{t-1} + \gamma_t(\tilde{\theta}_{t-1/2} - \tilde{\theta}_{t-1})$ \quad with $\gamma_t$ a step size
    }
    Output $\tilde{\theta}_T$
    \caption{IMSA with multiple imputation}\label{paper-3-alg:IMSA2}
\end{algorithm}

\begin{algorithm}[!htp] %[H]
    \SetAlgoNoLine
    Initialize $\hat{\theta}_0$ and $m$ copies of latent variables $u_{01},u_{02},...,u_{0m}$ \\
    \For{ $t = 0,1,2,..., T$}{
        \eIf{$t\leq T_0$}{
            \For{$j = 1,2, ..., m$}{
                Sample $u_{tj}$ given $(u_{(t-1)j},\hat{\theta}_{t-1})$ leaving $f(u|y;\hat{\theta}_{t-1})$ invariant by Algorithm~\ref{paper-3-alg:pMALA} \textbf{without preconditioning}
            }
           }{
            \For{$j = 1,2, ..., m$}{
                Sample $u_{tj}$ given $(u_{(t-1)j},\hat{\theta}_{t-1})$ leaving $f(u|y;\hat{\theta}_{t-1})$ invariant by Algorithm~\ref{paper-3-alg:pMALA} \textbf{with preconditioning}
            }
          }
        Compute $g = \frac{1}{m}\sum_{j = 1}^m \left\{ \frac{\partial}{\partial \theta}\log f(y, u_{tj};\theta)\mid_{\theta = \hat{\theta}_{t-1}}\right\}$\\
        Update $\hat{\theta}_t = \hat{\theta}_{t-1} + \gamma_t g$ \quad with $\gamma_t$ a step size
    }
    Output $\hat{\theta}_T$
    \caption{ScoreSA with multiple imputation}\label{paper-3-alg:ScoreSA2}
\end{algorithm}

\begin{algorithm}[!thp] %[H]
    \SetAlgoNoLine
    Input current latent variables $u$ and parameter $\theta$\\
    \eIf{Using Preconditioning}{
        Compute $\Sigma = [\nabla^2 Q(0)]^{-1}$\qquad\qquad  \#Evaluate inverse hessian at $u = 0$
    }{
        Set $\Sigma = I$
    }
    Initialize $u_{old} = u$ and $\xi = u - \frac{\epsilon^2}{2} \Sigma \nabla Q(u)$ with $\epsilon$ a step size\\

    \For{$i = 1,...,N$}{
        Sample $Z\sim N(0,\Sigma)$ and $w\sim \text{Uniform}(0,1)$ \\
        Compute $u^* = \xi + \epsilon Z$ \\
        Compute $\xi^* = u^* - \frac{\epsilon^2}{2} \Sigma \nabla Q(u^*)$ \\
        Compute $\rho = \exp\left\{Q(u_{old}) - Q(u^*) + \frac{1}{2\epsilon^2}(u^* - \xi)^{\T} \Sigma^{-1} (u^* - \xi) - \frac{1}{2\epsilon^2}(u_{old} - \xi^*)^{\T} \Sigma^{-1} (u_{old} - \xi^*) \right\}$
        \If{$w < \min(1,\rho)$}{
            Set $u_{old} = u^*, \xi = \xi^*$  \qquad\qquad \#Accept
        }
    }
    Set $u_{new} = u_{old}$ \\
    Output $u_{new}$
    \caption{MALA/pMALA for sampling from $f(u|y;\theta)$}\label{paper-3-alg:pMALA}
\end{algorithm}
\newpage

\begin{figure}[H]
    \centering
    \includegraphics[width = 0.8\textwidth]{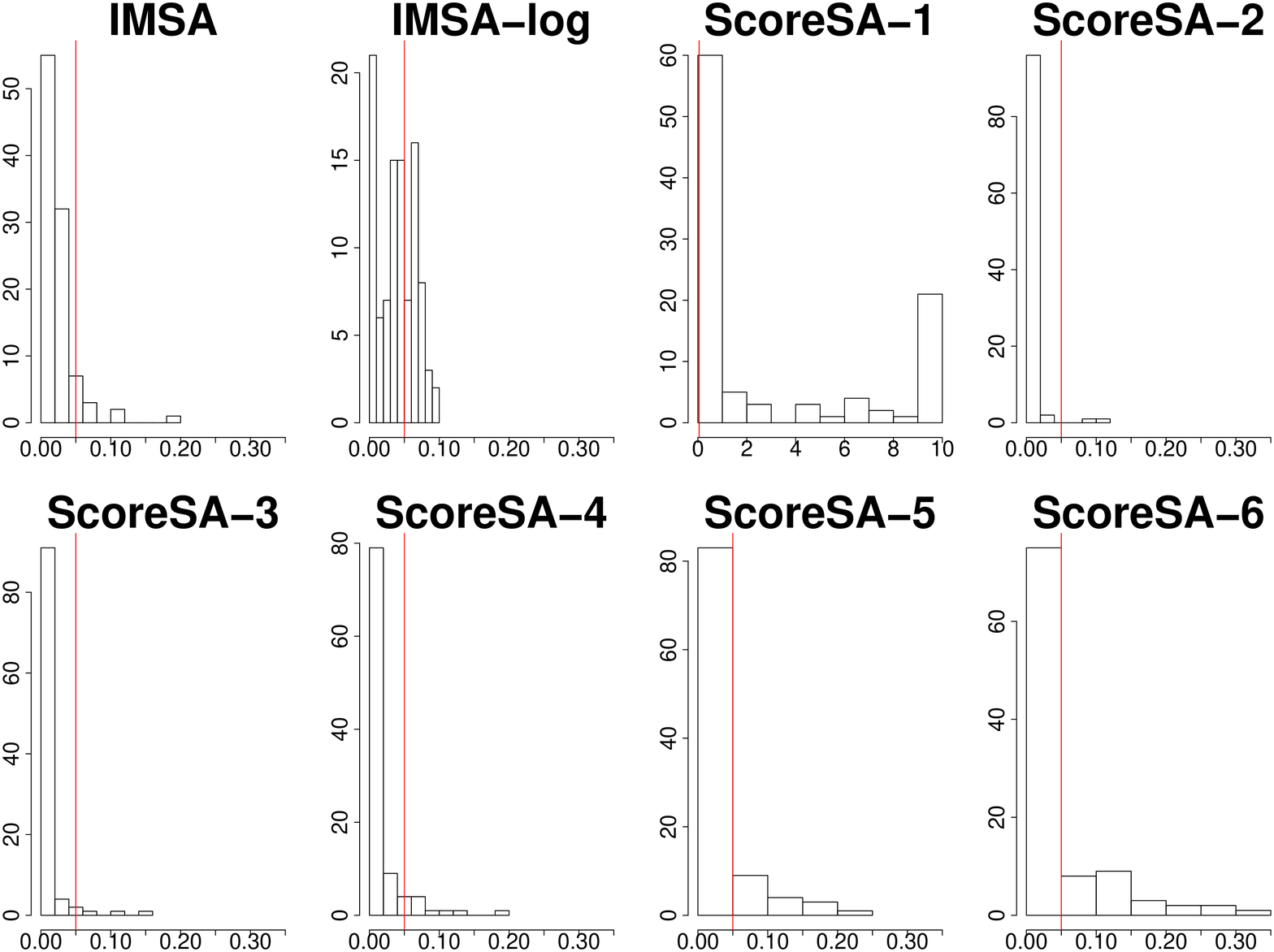}
    \caption{Histograms of rolling average of gradient norms in Booth--Hobert example}
    \label{paper-3-fig:logithist}
\end{figure}

\begin{figure}[H]
    \centering
    \includegraphics[width = 0.8\textwidth]{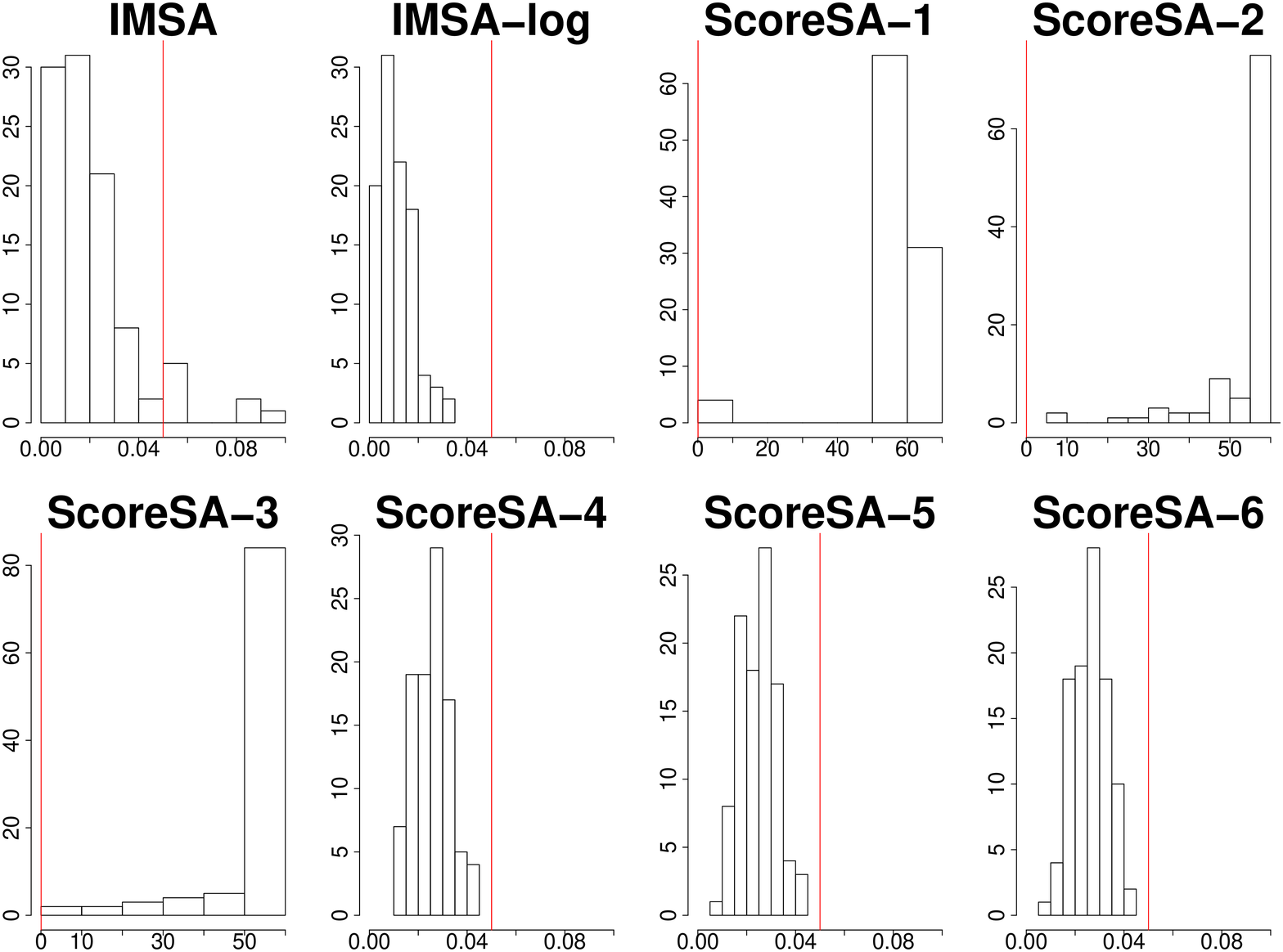}
    \caption{Histograms of rolling average of gradient norms in the salamander mating model}
    \label{paper-3-fig:salhist}
\end{figure}

The term $(\tilde{\theta}_{t-1/2} - \tilde{\theta}_{t-1})$ in IMSA serves a similar role as the gradient $\frac{\partial}{\partial \theta}\log f(y, u_{t};\theta)$ in ScoreSA. As
a way to monitor the converge, we record the norms $||\tilde{\theta}_{t-1/2} - \tilde{\theta}_{t-1}||_{\infty}$ for IMSA and $||\frac{\partial}{\partial \theta}\log f(y, u_{t};\theta)||_{\infty}$ for ScoreSA
at each iteration. We then compute rolling averages of the norms using a window length of $250$. Histograms of the minimum of the rolling averages over repeated runs are plotted in Figure~\ref{paper-3-fig:logithist}
and Figure~\ref{paper-3-fig:salhist} with red lines marking a threshold value of $.05$. A higher frequency below the threshold indicates better convergence
by a certain number of iterations. According to the plots,
in Booth--Hobert example all methods except for ScoreSA-1 show good convergence within 2000 iterations; in the salamander mating model, ScoreSA-1/2/3 do not converge as their $\sigma^2$ estimates explode,
while the remaining methods converge within $4000$ iterations.

\section*{Appendix B: IMSA with variance estimation}
IMSA with variance estimation of $\tilde{\theta}_{IMSA}$ is given in Algorithm~\ref{paper-3-alg:IMSA3}.

\begin{algorithm}[!tph] %[H]
    \SetAlgoNoLine
    Initialize $\tilde{\theta}_0$, $u_0$, and $(p+K)$-dimensional vector $\bar{r}_0$, $(p+K)\times(p+K)$ matrices $\bar{R}_0, \bar{H}_0$\\
    \For{ $t = 1,2, ..., T$}{
        Sample $u_t$ by MCMC given $u_{t-1}$ leaving $ f(u|y;\tilde{\theta}_{t-1})$ invariant
        Compute $\tilde{\theta}_{t-1/2} = \argmax_{\theta}\log f(y,u_t;\theta)$
        Compute $r = \tilde{\theta}_{t-1/2} - \tilde{\theta}_{t-1}$
        Compute $H =  - \frac{\partial^2}{\partial \theta \partial \theta^{\T}} \log f(y, u_t;\theta)\mid_{\theta = \tilde{\theta}_{t-1}} $
        Update $\bar{r}_t = \bar{r}_{t-1} + \rho_t (r - \bar{r}_{t-1})$ \quad  with $\rho_t$ a step size
        Update $\bar{R}_t = \bar{R}_{t-1} + \lambda_t ((1-\lambda_t)[r - \bar{r}_{t-1}][r - \bar{r}_{t-1}]^{\T} - \bar{R}_{t-1})$ \quad  with $\lambda_t$ a step size
        Update $\bar{H}_t = \bar{H}_{t-1} + \nu_t (H - \bar{H}_{t-1})$ \quad  with $\nu_t$ a step size
        Update $\tilde{\theta}_t = \tilde{\theta}_{t-1} + \gamma_t\,r$ \quad  with $\gamma_t$ a step size
    }
    Output $\tilde{\theta}_t$ as point estimate and $[\bar{H}_t(I - \bar{R}_t\bar{H}_t )]^{-1}$ as variance estimate
    \caption{IMSA with variance estimation}\label{paper-3-alg:IMSA3}
\end{algorithm}

\end{document}